\documentclass[12pt,showpacs,showkeys]{revtex4} 

 \usepackage{graphicx}

\usepackage{amssymb}
\usepackage{amsmath}

\def\ProgramName{FitSuite}
\def\LongProgramName{FitSuite}
\def\Mossbauer{M\"{o}ssbauer}

\newcommand{\dolt}[1]{\textit{#1}}
\newcommand{\doltSpec}[1]{\textit{#1}\-}
\newcommand{\egyspec}[1]{#1\-}
\newcommand{\vast}[1]{\textbf{#1}}

\newcommand{\ttinf}[1]{\text{#1}}
\begin{document}




\title{FitSuite a general program for simultaneous fitting (and simulation) of experimental data.}

\author{Szil\'{a}rd Sajti}
\email{szilard@rmki.kfki.hu}
\affiliation{KFKI Research Institute for Particle and Nuclear Physics, P.O. Box 49, H-1525 Budapest, Hungary}
\author{L\'{a}szl\'{o} De\'{a}k}\affiliation{KFKI Research Institute for Particle and Nuclear Physics, P.O. Box 49, H-1525 Budapest, Hungary}
\author{L\'{a}szl\'{o} Botty\'{a}n}\affiliation{KFKI Research Institute for Particle and Nuclear Physics, P.O. Box 49, H-1525 Budapest, Hungary}



\begin{abstract}
In order to get accurate information about complex systems depending on a lot of parameters, frequently different experimental methods and/or different experimental conditions are used. The evaluation of these data sets is quite often a problem. The correct approach is the simultaneous fitting, which is rarely used, because only a very few programs are using it and even those cover usually a narrow field of physics. \ProgramName{} was written to tackle this problem, by providing a general and extendable environment for simultaneous fitting and simulation. Currently it is used for M\"{o}ssbauer spectroscopy, grazing incidence neutron and (non)resonant X-ray reflectometry, but in principle other experimental methods can also be added.
\end{abstract}


\maketitle
\section{Introduction}\label{sec:Intro}
Nowadays scientists examine more and more complex systems, which depend on a lot of parameters. To get a correct, accurate and detailed picture about the processes and phenomena in these systems we need more and more data. These can be obtained by measurements performed on the same sample with different experimental methods, which may be sensitive for different parameters, and/or with the same method performed using slightly different experimental conditions, such as temperature, pressure, magnetic field, etc. Such data often depend partly on the same set of sample and experimental parameters, therefore a simultaneous evaluation of all the data is prerequisite. However, data evaluation programs are dominantly organized around a single experimental method and a single theoretical approach used for simulation of this problem, therefore a simultaneous access to the data for a common fitting algorithm is not typical. Lacking suitable programs for simultaneous data evaluation, experimentalist determine some of the parameters from one kind of measurement, assume them error free and keep them constant when evaluating other experiments, which is obviously an incorrect approach. Besides, for different problems different programs are used, which makes it very difficult to tune the parameters of such problems and their errors and correlations to each other and to extend or modify the `codes' used to simulate the results of different experimental methods, which in science is a frequently arising problem.

There are some programs which are able to perform simultaneous fitting, but usually they are restricted to a `narrow' field of physics, as e.g. IFEFFIT \cite{IFEFFIT} to X(-ray)A(bsorption)F(ine)S(tructure), RefFIT \cite{RefFIT} used to analyze optical spectra of solids, SimulReflec \cite{SIMULREFLEC} for neutron and X-ray reflectometry. They were written having in mind the specific problems, and the requirements demanded by them. Even if they are written in a way, that makes possible further extensions, generalizations; they are not apt to include a problem from a quite different field, without writting essentially a new program. Doing this we have to spend a lot of time with miscellaneous problems having nothing to do with physics, in order to have just a feasible user interface.

Over the past years, Hartmut Spiering has developed the general and versatile data fitting environment EFFI (Environment For FItting) \cite{Spiering00}, which aimed to solve these problems, and which has been very efficiently applied for the evaluation of many sets of ‘conventional’ transmission and ‘synchrotron’ \Mossbauer{} spectra the latter including grazing-incidence, i.e., synchrotron \Mossbauer{} reflectometry (SMR) \cite{NDL00} measurements, both time-differential and time-integral. The main and yet essential disadvantage of this program was the fact that its user interface was written using development tools available in the early eighties, and therefore, in spite of its scientific merits, its user-friendliness was considerably limited. Another problem is the lack of documentation, which hinders effectively its development furher beyond a certain degree.

Therefore it was targeted to write a new thoroughly documented program, called \LongProgramName{}, with a graphical user interface, written in C++, retaining all the good ideas, principles available in EFFI, but rethought in order to generalize, extend them where it is possible.

In this article we are presenting the test version of \ProgramName{}, freely downloadable from the home page of the program \cite{EffiWebPage} with the only provision of properly acknowledging its usage in upcoming publications.
Presently, it is capable of simultaneously fitting several data sets of the following kinds of experiments:
\begin{itemize}
\item Conventional \Mossbauer{} absorption and emission spectroscopy
\item X-ray reflectometry
\item Nuclear resonant forward scattering of synchrotron radiation: time differential mode
\item Nuclear resonant forward scattering of synchrotron radiation: stroboscopic mode
\item Synchrotron \Mossbauer{} reflectometry: time integral, time differential and stroboscopic modes
\item Specular polarized neutron reflectometry
\item Off-specular polarized neutron reflectometry.
\end{itemize}
The addition of new kinds of experiments is possible.

\section{Considerations, goals}
Writting \ProgramName{} we had several considerations, which the program should satisfy. These will be summarized shortly in this section.

 The program should provide a general abstract interface for simultaneous fitting and/or simulation of different experimental methods, in order to be able to add new type of problems with minimal effort, without changing the program itself. There should be an interface for the rare users, who want to add a new type of experimental method, giving just the functions, subroutines needed for simulations, and some description of the parameters and the concepts used in the modelled system (e.g.: sample, detector, source, layer). As the addition of new methods should be possible without recompiling the whole program, modularity is needed. For the goals of the program the object oriented language C++ seemed to be the most appropriate. But as, there are a lot of codes available in Fortran, and sorrily there are people, who do not like to learn new program languages, it was an additional requirement to be possible to write the functions (subroutines) not only in and/or C(++), but also in Fortran.

There is another type of user (most of them), who just want to use the `codes' provided by others in order to evaluate their experimental results or simulate their problems. They need another interface, in order to be able to use the program easily, with minimal effort. The interface should be a graphical user interface (GUI), but the program `core' should be separated from the GUI. This is needed for several reasons, which are connected with further possible plans about the extension of the features of the current program. Sometimes a console interface can be more useful, than a GUI. If the user would like to run the program on a cluster or a grid, there is only one GUI needed. The change of the GUI will be easier, if it is separated from the core.

There was another requirement to use only packages, which make possible to compile the program for different platforms (primarily Linux and Windows) without much pain. Therefore we chose the Qt package from Trolltech for the GUI development.

In the following we will try to summarize, what are the requirements to describe an experimental method and its subject in an abstract way. This may seem to be quite easy, as we usually are not aware the concepts we use without hesitation and much thinking, describing or calculating problems related to a physical system.

We will use a few concepts used in C++ and every object oriented language. These will be concerned very slightly, therefore we hope that it will not cause problems, even if the reader is not acquainted with them. If there is a need of better understanding, we recommend any book related to these languages (or just a fast search on the internet), and skip the parts in parenthesis boldly.

\section{Basic concepts}\label{sec:BaseConc}
First we just sketch the main concepts used by \ProgramName{} and their relations to each other and we sunk into the details only thereafter. In \ProgramName{} we have always a \vast{simultaneous fit project} (represented by the class CLSimultanFitProject) which is consisted of \vast{fitting problems} which the user would like to fit simultaneously (represented by classes CLGenFitProblem and CLFitProblem). A fitting problem is consisted of the \vast{experimental data} (represented by class CLExperimentalData) and of the computer \vast{model} of the experiment (represented by class CLModel). In the following we will see in details, what a model is, what is it consisted of. We will get into the details only to such a depth, which may be useful for a user, who would like to add new problem types to the program.

\subsection{Model and its parts}\label{sec:ModelAndItsParts}
A model of an experiment contains the `sketch' of the experimental setup and the system under study, as a physical system and the algorithms with which the experiment can be simulated, its results can be calculated. Before this text would start to get too complex and not too understandable, let see an ordinary example by which we can explain what a model is in \ProgramName{} more smoothly. Let assume that someone has a model describing an experiment (or rather models of experiments, if we want simultaneous fits) with a body, and try to answer the questions: How should this model look like and what concepts it needs? First we try to forget the experimental setup just concentrate on the subject of the experiment, i.e. on the body. Clearly we cannot do this perfectly, as the model of the body will depend very strongly on the experimental circumstances. E.g. in a very simple throwing experiment in rare atmosphere the body can be conceived just as a particle having mass. But we need a more detailed model in dense atmosphere and to complicate it further we can allow the body to change its shape. In these cases we have to know more about the structure, the building blocks of the body which can influence its drag coefficient (air resistance) and the parameters with which these structural elements can be characterized. In \ProgramName{} these parameters are called \vast{properties} (represented by class CLProperty) and the structural elements are called \vast{physical objects} or \vast{physical notions} (represented by the class CLPhysObjNot the name is created by putting together \vast{obj}ect and \vast{not}ion). The name \dolt{physical notion} is used because in some cases the noun {\it object} is not appropriate (E.g. stating about an object that it is consisted of a specific type of a material, it is convenient to describe the material type with the same CLPhysObjNot class as the physical objects, in spite of the fact that it cannot be called an object and may not be a property as it may contain physical objects, as characteristic atoms, molecule groups, in \Mossbauer{} spectroscopy sites. Naturally, we could define another class for notions, or properties which may contain physical objects, but these ways would lead to a more complicated program structure.) and I did not wanted to use the word \dolt{concept}. In the following, in order to be short, we will write about physical object even if it is a notion.

\begin{figure}
\includegraphics[clip,width=125mm]{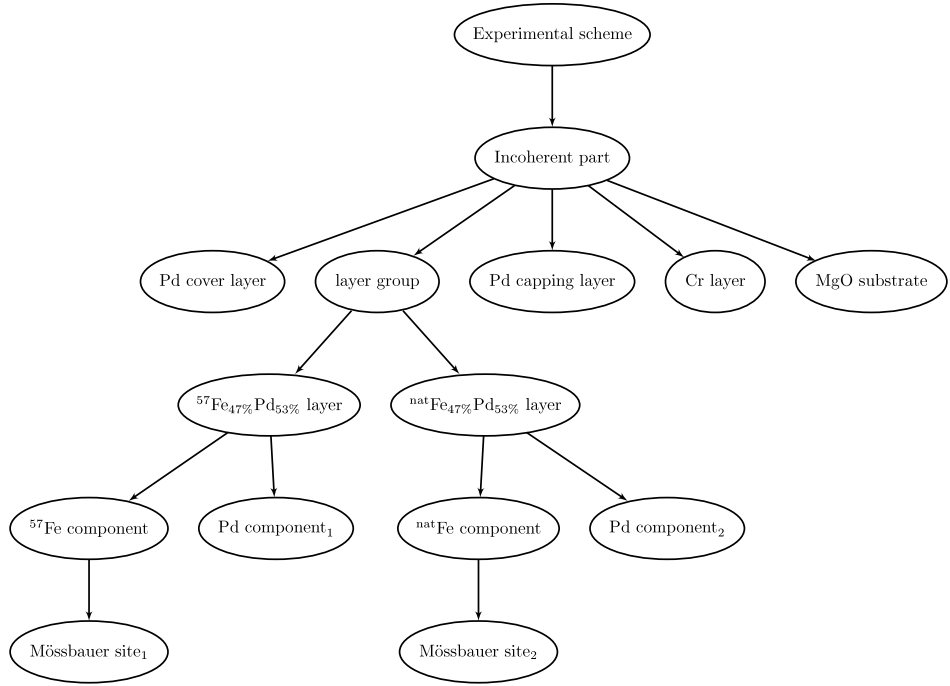} 
\caption{An example of the tree structure belonging to an isotopic multilayer system used to examine the self-diffusion of iron atoms in FePd alloys \cite{DaniCikk}.}\label{figureTree}
\end{figure}

Thus we have now the subject of the experiment as a \vast{physical object}, which is built up from other \vast{physical objects} and which are characterized by \vast{properties}. We can fix without making constraints on the generality, that the hierarchy of the physical objects should have a simple tree structure (there is one root object containing everything in the hierarchy, and there are no loops in the corresponding graph). We want to describe not only the subject of the experiment but the whole experimental setup, so we can have a similar description of the experimental apparatus and the environment in which the experiment is performed, but everything detailed only to an extent ensuring that the problem can be simulated without having too much unnecessary parameters. (It is not hard to see, that the requirement of having only the necessary parameters would be too strict.) So we can fix that the \vast{main (or root) object} of the extended tree of physical objects should be always the \vast{experimental scheme} which contains the parts of the experimental setup and of the system under study, as it can be seen in example shown in Fig. \ref{figureTree}. (I am not sure that this is the best word for this concept: experimental-setup, -world, -universe, -system were also among candidates, but each may be mixed with something else). This main object (CLPhysObjNot::MainObj) is contained by the model (CLModel) and all the other physical objects are  `children' (CLPhysObjNot::ChildrenList) or `descendants' (CLPhysObjNot::DescendantsList) of this object.

Some of the \vast{physical objects} may have their own models, which makes possible to perform some simulations of the physical subsystem described by these objects and the results of these simulations may be used by a model belonging to an object containing these objects as children or descendants.

\subsection{Prototypes}\label{sec:Prototypes}
If we want to give the possibility to the user to choose the subject (e.g. not only two-winged but three-winged bodies also) of his or her experiment (and the experimental setup also) flexible, but of course only within certain limits, we have to (at least we went into this way to tackle this problem) define \vast{prototypes of models, physical objects and properties} (represented by classes CLProtoModel, CLProtoPhysObjNot and CLProtoProperty respectively). The relation of a model prototype object to the corresponding model objects (here we use the word `objects' in programming technique meaning) is similar to the relation between the set of building block types plus the knowledge of the connection possibilities and the `maquettes' built up using these blocks and rules (and not what is implied by the word prototype).

Now let see what informations these prototypes should contain, how they should look like. First of all, we need something to identify, differentiate them. In \ProgramName{} we use for this purpose names and integer numbers. The first is for humans, the second is generated internally and is used by the simulation functions, in which the programmer can refer to them by variable names generated from the above mentioned names (as the programmer is human).

Here arises the question, that when should we regard two prototypes different. At the level of model prototypes there is no problem, they all have to have different names at least the ones which can have a role in the same simultaneous fit project. Therefore (but not only because of this) the model prototypes are stored in repositories  (represented by CLProtoModelRepository). In a simultaneous fit project we can use only the model types of one repository (at least in current version of \ProgramName{}).

At the level of physical object prototypes we require uniqueness only within the model prototype to which they belong. At the level of property prototypes we require uniqueness only within the prototype of physical object to which it belongs (e.g. the sample and the domains, or other miniature structures on it also may have diameter, although their have different value and order of magnitude). We have to know that to which object type a certain property belongs, that is why we have the whole hierarchy of physical objects. In the same way we have to know the parents (grand-...-grand-parents) of an object (e.g. the body may have screws on its wings and fuselage as well). We have to know that an object of a specific type which type of objects may and how many may (or have to) contain or is there a limit at all. With these pieces of information we can help the user when she builds up the model of her experimental scheme. We may allow only the appropriate combination of building blocks (or we can warn the user). Furthermore we may be able to select the parameters, properties, objects, which we are interested in according to complex type criteria (CLTypeSelectionCriterion, CLSelection) given by us.

The tree of the physical objects is an ordered structure by the parent-child relations `vertically' and is also ordered `horizontally', as the list classes (from standard C++ library), which are used for the storage of children of physical objects can be conceived as an array with beginning and end into (from) which we can insert (remove) elements at (from) arbitrary position. Sometimes, this order is a requirement (e.g. thin layer systems) sometimes is not, but even in the second case it is convenient. In the prototypes of physical objects is also an order, the possible parent types and child types are also ordered. therefore in the children list of a physical object the sequences of different type of objects also have strict orders. (E.g. it is not a requirement, but it is logical to have the order: source(s), sample(s), detector(s) in a scattering, absorption, etc. experiment.) The properties also have a strict order within a physical object, which is determined by the order of their types in the corresponding physical object type. Furthermore, because of the availability of these orders, we may have the physical objects of the same type of a model in an ordered list (CLProtoPhysObjNot::RepresentativesList). This may be convenient when we want to perform some operation on the same type of objects and their properties without going through the tree hierarchy.

\subsection{Group}\label{sec:Groups}
In an ordered system as a layer structure, we may have periodic sequences. For description of such sequences we use (as for thin layer structures is usual) \vast{groups}. In \ProgramName{} a group is a physical object, which contains physical objects of the same type that it belongs to, but it has no properties and its repetition number (CLPhysObjNot::Nrep) is greater than 0. In the physical object type we can specify whether that type may have group or not and how deep these groups may be embedded into each other (CLProtoPhysObjNot::GroupDepth).

\subsection{Property}\label{sec:Property}
Above, we just mentioned the concept of property, but did not examined it in details or the requirements arising with it.
The properties are (C++) objects representing physical quantities and other numbers, which are needed in order to simulate the problem properly, mainly arising because the calculations are performed by a computer and because the experimental results are always discrete data sets.

First let see what is needed in order to represent physical quantities. A physical quantity has an algebraic structure. Even though all components of a nonscalar physical quantity could be represented by independent scalars, sometimes it may be convenient to know that these components belong together. E.g. when the user lists out all the components of such a quantity it is good to give only the name of the quantity and not all of the components. The algebraic structure in computer representation should not be always identical with the mathematical structure used in science. E.g. the components of a symmetric tensor can be represented by a vector (it would be more appropriate to call it an array) and not by a matrix. In current \ProgramName{} the E(lectric)F(ield)G(radient) tensor is represented as a 5 element vector. Three of them are the Euler angles giving the orientation of the coordinate system in which the EFG can be given by the remaining two parameters. In this case, it would be more appropriate to speak about parameters determining the tensor, than components, but we do not want to introduce a new concept just because of this.

So the properties have (algebraic) structures, and are built up from scalar components. Each component may have its unique name (within the property, and if the user did not name it, the program will generate names using ordinal numbers), its value, its minimum-, maximum value, order of magnitude. To a physical quantity naturally belongs some unit. In case of a nonscalar quantity the different components may be measured in different units (CLProtoProperty::DefaultUnits). E.g. the magnetic induction vector given in spherical coordinates has a radial component \(B_r\) in Tesla (or Gauss) and the angles \(B_\vartheta\), \(B_\varphi\) in degree (or radian). As it was mentioned above the properties may represent numbers which are not physical quantities. This means not in all cases that this type of property has no physical significance at all. E.g. at the moment, symmetries of the sites are represented by three integer numbers, \(C_{nzn}\) is one of them, it determines whether the axis \(z\) is a symmetry axis or not, and if it is how many fold this rotation symmetry is. Some numbers could be used as switches. E.g.:. But the properties can also represent numbers which do not have real physical significance. These typically just give an arbitrary size of an array, which can say something about the (sampling frequency) resolution with which the simulation or experiment was performed at most.

To each component belongs an integer number, which we call logical bit collections (more specifically an enumeration type named EnLogicalCollection) whose bits contain information specifying further the role of the corresponding component in the model.  E.g. whether they are constant, independent variables, internal variable (handled internally during simulation or fit and invisible for the user), free or fix; whether they were changed since the last iteration step, etc. 

To each property type belongs some help contained by a string (CLProtoProperty::Notes) and an url reference (CLProtoProperty::UrlFragment) to the place, where a more detailed help may be available.

The models, physical objects and properties have to contain some information according to which their prototype can be determined. (This is solved in all cases with the help of a pointer named \dolt{Proto} namely CLsubProperty::Proto, CLPhysObjNot::Proto, CLModel::Proto pointing to the proper prototype namely an object of class CLProtoProperty, CLProtoPhysObjNot and CLProtoModel, respectively).

\subsection{Parameter name convention}\label{sec:ParNameConv}
The property uses always the name of its prototype (CLProtoProperty::Name), as it is unique in the object (type), which is characterized by it. Therefore with the object name and the property type name  we can find it always. (E.g. thickness is always thickness we just say that it is the thickness of the body's first left or back right wing, or its n-th screw on its right tail wing upper part.) In case of physical objects, models the prototype name is not enough, they have to have their own names. But of course we can use the prototype names even in this case to generate an automatic name. Only those physical object names are allowed, which are unique within a main model (i.e. not a submodel). In a simultaneous fit project a property may be identified unequivocally by the model name (main model and no submodel), the physical object name and the property prototype name. As we will see later, from this object tree structure we will generate a parameter list, used during the (simultaneous) fit or simulation. Because of having these `constraints' on the names, each parameter belonging to a model can be and is identified using the name convention: \doltSpec{Model\-Name}\verb+=>+\doltSpec{Object\-Name}\verb+:>+\doltSpec{Property\-Name}::\doltSpec{Component\-Name} or in case of scalars just \doltSpec{Model\-Name}\verb+=>+\doltSpec{Object\-Name}\verb+:>+\doltSpec{Property\-Name}. In case of complex scalars we have automatic component names \dolt{.re} and \dolt{.im}. For complex vectors (and other nonscalars), the component names get \dolt{.re} and \dolt{.im} as an additional suffix. If we had no unique physical object name in a model, but only its parent, then we should give the whole hierarchy of object names (e.g.: \doltSpec{Model\-Name}\verb+=>+\doltSpec{Root\-Object\-Name}\verb+->+\doltSpec{Grand\dots Grand\-Parent\-Name}\verb+->+\dots\verb+->+\doltSpec{Parent\-Name}\verb+-+\verb+>+\doltSpec{Object\-Name}\verb+:>+\doltSpec{Property\-Name}::\doltSpec{Com\-pon\-ent\-Name}), which could be very long and quite impractical. Because of this parameter name convention and some others coming later on, the names should not contain the character sequences used as separators and suffixes: `\verb+=>+',  `\verb+:>+', `::', `.', `,', `\verb+*>+', `\verb+>>+', `.re', `.im'. Use of whitespaces should be avoided also, because it can cause bugs reading the simultaneous fit projects from files.

\subsection{Beyond the tree structure}\label{sec:BeyTreeStruc}
Sometimes we do not have `well defined' physical objects, but rather a statistical ensemble of them. In these cases we may need distributions and/or correlation functions. In \ProgramName{} presently, we have only correlation functions of 2-order, and even those only for a very specific case. Later on this should be rewritten if there is a requirement for it. 

In order to be more understandable, let see the above mentioned problem. There is a magnetic multilayer system. Some of the layers are consisted of magnetic domains of \( n \) type. The domains of different layers are antiferromagnetically coupled to each other. For description of off-specular resonant X-ray (\Mossbauer{}) reflection \cite{DeakOffsp} on such samples we have to know that: which layer, what type of domains is consisted of; what is the fraction of the \( i \)-th type of domain in the \( m \)-th layer. Besides this there are some correlation functions between the domains in different layers, e.g. \( c_{ik,jl}(\dots) \) between the \( i \)-th layer`s \( k \)-th type of domain and the \( j \)-th layer`s \( l \)-th type of domain. Each such correlation function has its own parameters, which we should be able to fit.

It is obvious, that such a problem cannot be handled with the tree structure shown before, as the correlation functions belong to two objects and it would be a waste to add to each layer the same domain types. Therefore we created classes to have symbolic objects also. This is also too specific currently and perhaps unsatisfactorily tested.  A symbolic physical object (CLSymbPhysObjNot) is similar to the symbolic links in the \dolt{Unix} file systems, or the application links in another well known operating system family, but there are a lot of differences. The symbolic objects also have prototypes. (represented by the class CLProtoSymbPhysObjNot. A CLProtoSymbPhysObjNot object has just a pointer CLProtoSymbPhysObjNot::ObjectType to the physical object prototype CLProtoPhysObjNot whose representatives may be symbolically linked as a child to objects, whose type is restricted by a list CLProtoSymbPhysObjNot::ParentTypes.) This way we can hinder the user creating symbolic links which would be meaningless, or could result program faults, i.e. this is a requirement of a `userproof' program. Besides this we may have additional constraints:
\begin{itemize}
\item Sometimes it may be useful to forbid to have some type of `brothers' (CLProtoSymbPhysObjNot::ExcludedBrothers) and properties in the parent (CLProtoSymbPhysObjNot::ExcludedProperties). In the first case we may not add this type of symbolic object to an object containing already such a child (listed among \dolt{ExcludedBrothers}), or if it contains already such a symbolic child, we cannot add any child, whose type is listed among \dolt{ExcludedBrothers}. In case of excluded properties, the properties are there, but we do not use them (it is planned to hide them from the user in the future versions), for this reason a logical bit (lcExcluded) are set to true  for each component of these properties, when we add such a symbolic child.
\item We may specify some properties also, which may be different, for each symbolic physical object, even if they are `links' to the same object. These properties we call overloaded properties, and they are given by a list (CLProtoSymbPhysObjNot::NamesOfOverloadedProperties) containing their names. In case of the off-specular example the fraction of domains in a layer, changes from layer to layer.
\end{itemize}

The correlation function (CLCorrelationFunction) is a bit similar to a physical object, it has properties and protototype (CLProtoCorrelationFunction), which contains a reference (function pointer) to the algorithm used for calculation of this function. (We have two special cases, the fully correlated and the totally uncorrelated case, when the value of the correlation function is identically 1 and 0, respectively. Therefore we have an enumeration type (CLCorrelationFunction::Correlated), according to which we can decide, that we have these two extreme cases, or we use a real function belonging to the corresponding prototype.) Its name is the identical with the name of its prototype (as e.g.: a Gauss or a Lorentz-function is always the same, just its parameter values may be different). The parameters belonging to a symbolic physical object and for a correlation function, obviously should be different from what we have shown earlier. For the first one we have \doltSpec{Model\-Name}\verb+=>+\doltSpec{Object\-Name}\verb+*>+\doltSpec{Symbolic\-Child\-Object\-Name}\verb+:>+\doltSpec{Property\-Name}::\doltSpec{Com\-pon\-ent\-Name} (e.g. \egyspec{Model\-X}\verb+=>+\egyspec{nth\-Layer}\verb+*>+\egyspec{domain\-Up}\verb+:>+\egyspec{size}), and for the second one \doltSpec{Model\-Name}\verb+=>+\doltSpec{Object\-Name\_1}\verb+*>+\doltSpec{Symbolic\-Child\-Object\-Name\_1}\verb+,+\doltSpec{Object\-Name\_2}\verb+*>+\doltSpec{Sym\-bol\-ic\-Child\-Ob\-ject\-Name\_2}\verb+>>+\doltSpec{Function\-Name}\verb+:>+\doltSpec{Property\-Name}::\doltSpec{Com\-pon\-ent\-Name} (e.g. \egyspec{Model\-X}\verb+=>+\egyspec{\-nth\-Layer}\verb+*>+\egyspec{domain\-Up,mth\-Layer}\verb+*>+\egyspec{domain\-Down}\verb+>>+\egyspec{Lorentz}\verb+:>+\egyspec{half\-Width}).

\subsection{Plotting, independent variables for simulation}\label{sec:PlotIndepVar}
To have the results of a simulation or fit in an appropriate way we have to plot the results, therefore we have to give some information for the computer, what sort of plot we need: as the scaling (e.g. logarithmic or linear), the labels of the axes, which arrays contain the results, which properties determine the array sizes, etc. For this aim we use also a class (CLPlotType). Each model prototype contains a list (CLProtoModel::PlotTypes) of them, from which the user can choose, if (s)he would like to. 

If we do not have data, but we would like to simulate, we have to tell to the computer, for which independent variable values should be the simulation performed. There may be several types of these also, depending on what is the independent variable (e.g. in case of neutron reflectometry, the wavelength, the scattering wavevector, the angle of incidence, etc.), which may be a scalar or a vector independent variable. Besides this, even if we have data, we should know what is there the independent variable. For this we have also a class (CLSimulationPointsGenerator). This contains:
\begin{itemize}
\item the names of the properties and their components, which determine the range and the distribution of the independent variables, i.e. their values;
\item the names of related properties, which are used only for the generation of the independent variables, and which should be hidden from the user, when (s)he uses another type of simulation point generator, which does not depend on them;
\item the name of the property, in which we store the type identification number belonging to the simulation point generator. This is needed, as in the simulation functions (subroutines) we have to know, what should be calculated.
\end{itemize}
(Sometimes just the conversion of the independent variables could be enough, but not always, this class is for that cases. The independent variable conversion could be an additional step.)
\subsection{Transformation matrix technique, parameter list generation}\label{sec:TransMatrPar}
The `optimization' methods used for fitting require a parameter vector and not an object tree structure with properties. Furthermore in case of simultaneous fitting we usually have the results of experiments performed in a bit different environment (external field, temperature, etc.) and/or different type of experiments using the same `sample'. Therefore there is a lot of common parameters. To eliminate this type of redundancy and as it is also convenient for the user to use as few parameters as possible (as it is more transparent for human and easier to fit in a parameter space with lower dimension at least if we want to get correct results) transformation matrix technique is used \cite{Kulcsar71}. For this we need also parameter vector (array). Because of these considerations we have to generate the parameter vector and the initial transformation matrix from the object tree structure. The model parameters which still contain all the redundancy can be collected in an array \( \mathbf{p}=\left(\mathbf{p_1},\mathbf{p_2},\dots,\mathbf{p_n}\right),\) where \( \mathbf{p_i} \) is the array containing all the parameters belonging to the \(i\)-th model in the current simultaneous fit project. Let denote the array of the fitting (or if you like simulation) parameters with \( \mathbf{P} \) and the transformation matrix with \( \underline{\underline{\mathrm{T}}}. \) The transformation matrix technique uses the expression \(\mathbf{p}=\underline{\underline{\mathrm{T}}}\mathbf{P}, \) where \(\dim \mathbf{P} \leqslant \dim \mathbf{p}\). Above was mentioned that this technique is used in order to eliminate the redundancy arising because of the common parameters, but this is not the unique reason. We can take into account some possible linear relations between the parameters also, which also is a redundancy of course. Furthermore we could generalize this technique using some additional nonlinear transformation operator. In that case we would have \( \mathbf{p}=\underline{\underline{\mathrm{T}}}\mathbf{P}+\mathbf{\mathrm{NL}}(\mathbf{P}) \). (The components of the nonlinear operator could be function pointers, given by the user as an assembly like code, or using some mathematical parser package as \dolt{muParser} and \dolt{MTParser}. The inhomogeneous transformation can be useful sometimes also. Sorrily this is not available on the level of GUI either in present program.)

Now arises the question, how to generate the initial \( \underline{\underline{\mathrm{T}}}\) matrix and the arrays  \( \mathbf{P} \) and \(\mathbf{p} \), which the user can change on the GUI according his ideas. It is advisable to take into account that there are parameters which according to expectations will not have interdependencies and therefore the \( \underline{\underline{\mathrm{T}}}\) matrix can be `block diagonalized'. It is more transparent to handle submatrices with lower dimensions, than one extended sparse matrix. Therefore we have to categorize the parameters according to our expectation whether the subspace stretched by a subset of them may have interdependencies or this is very unlikely. (If the user finds a case, where our expectations are not met, (s)he is able to unite or split the submatrices, thus our choice here is not a constraint.) The initial submatrices generally are identity matrices, but not always. E.g. the thickness of a multilayer sample will be the sum of the layer thicknesses; in \Mossbauer{} spectroscopy in a doublet site, the line positions and the measure of the splitting and the isomer shift will not be independent, etc.

A model parameter type (e.g. layer thickness, magnitude of the external magnetic field on the sample, etc.) can be specified by the physical object types, by the property type and the property component. In a general case we are able to differentiate not only by the object type, but by the branch of objects starting from the root object in the given model type, we can take into account this way the parents, grand...grandparents, the `pedigree' of the object. (E.g. in the throwing example, the length (material) of the screws on the wings and on the fuselage can be quite different, and can be regarded independent from each other.) They may belong to quite different subspace, category. Each model has a partition (CLPartition ) class which contains a category (CLCategory) list. Each parameter type belongs to only one category. To each category belongs an initial transformation matrix, which is used during the generation of the initial transformation submatrix belonging to the category. The real and integer parameters are handled separately as we do not want to guard the parameters against conversion (rounding) errors, and the integer parameters are never fitted. Therefore the real and integer parameters have separate partitions  (CLProtoModel::Partition and CLProtoModel::intPartition, respectively) and of course separate arrays and transformation matrices.
\subsection{Arrays and algorithms}\label{sec:ArrAlg}
In order to simulate we have to provide the algorithms for the model also. To each model belongs three function (in Fortran subroutine). One is used for the simulation. In the simulation we use arrays, containing the spectra, intermediate results, auxiliary arrays. These arrays, at least some of them should be initialized with values different from 0, before the first simulation of the fit iteration. For this we have another function. It is clear, that we have to give the size of these arrays also. Therefore the array initialization should be preceded by the array size initialization. The third function is used for this. Later we will see how this functions should look like, how we can write such one. We classify the arrays according to their roles into five main groups:
\begin{itemize}
\item The input arrays are initialized before the first iteration step.
\item The output arrays are (usually) set to 0 initially.
\item The variable auxiliary arrays are used internally, usually are set to 0 initially.
\item The constant auxiliary arrays are set only before the first iteration step, and not changed thereafter.
\item The constant integer auxiliary arrays are set also only before the first iteration step, and not changed thereafter.
\end{itemize}

Thus we have parameter arrays and transformation matrices, and the simulation and the two initialization functions, but we are still not done, as during the simulation we may need the structure also. We have to provide this information for the simulation functions someway. If we would use only C(++) language we could use the CLPhysObjNot objects or something similar, e.g. generated structures embedded into each other using (void* or just) pointers. But sorrily we use also Fortran, where we cannot embed structures into each other. (Even Fortran versions later than 77 - sorry Fortran believers - are childish, a joke in this regard.) Therefore we use the `information array' pinf (CLModel::pinf) generated by the program. In the following we will not go into the structure of this array, as it is quite complex, it can be found in the program documentation, and to write the three type of function needed for simulation and initialization it is enough to know the auxiliary functions, some of which are shown in the appendix \ref{sec:appA}.
\subsection{Following changes}\label{sec:FollCh}
During an iteration (fit) it is useful to calculate only if necessary. If an auxiliary array was calculated in the former iteration step, and the parameters and arrays on which it depends were not changed, there is no reason to calculate it again, especially if it takes a lot of computation time. For this reason we have functions to follow, the changes of the component of the properties and arrays.
The changes have three sources:
\begin{itemize}
\item[-]change of the parameters, input arrays by the user, this happens always before starting an iteration or simulation;
\item[-]change by the fitting method, this happens between the simulations, only the free parameters and the arrays may be changed this way;
\item[-]change of internal variables, this is done by the simulation and initialization functions, thus this is essentially the problem of the code writer.
\end{itemize}
(Structural changes, as removing or adding a layer, would be another class of changes, but that would lead us too far away, and the handling of this problem is out of our plans in near future.)
Therefore we have to know, whether was a simulation (or iteration) run before the current calculation, had been there an user interaction since then, are there free parameters, is the current function call the first one during the fit.
When the user changes a parameter in the user interface, or changes an input array, the proper bit (lcChanged) of the corresponding logical bit collection is set to true. Similarly in case of free parameters another bit (lcFree) is set to true.
The initialization and simulation functions have logical arguments, determining whether the function was already called, or is the first call during an iteration.
Using these and some auxiliary functions (not shown in this article), and proper coding, we can decide when we can jump same code parts during simulation or fit, as there was no change. 

Besides the auxiliary functions needed to write simulation and initialization functions, we also have to create the model types for the program.  In future for this task we will use another program with graphical user interface, in order to decrease the number of possible errors, in this process. But now we have to write a C++ program, as it can be seen in the appendix \ref{sec:appB} this can be done quite mechanically.
\section{The user interface of \ProgramName{}}

If you know already the principles used in \ProgramName{}, which we have shown in the former section, there is not much to know about the user interface. Therefore we just skim over it shortly.
Starting the program, the user can start a new project or load a previously saved one. We can save our project anytime. For building up the object tree structure of the models, we use an interface similar to the treeviews used every day to handle our file system. The main difference is, that here instead of directories we have physical objects, and instead of files properties. Furthermore we are constrained by the rules given by the model type. The data sets, parameters and transformation matrices, have their own editors, using `spreadsheets'. For plotting the package Qwt is used, gnuplot files are also generated.

For fitting several methods are available. Most of them is a slightly modified version of the optimization functions available in Numerical Recipes \cite{NumRec}. During fitting, we optimize always the \(\chi^2\) in current version. Later this may be changed. Confidence limits, covariance matrix of the free fitting parameters are calculated after fitting was finished.

Further details can be found in the User Manual and in the demos available at the homepage of the program.

\begin{figure}
\includegraphics[clip,width=125mm]{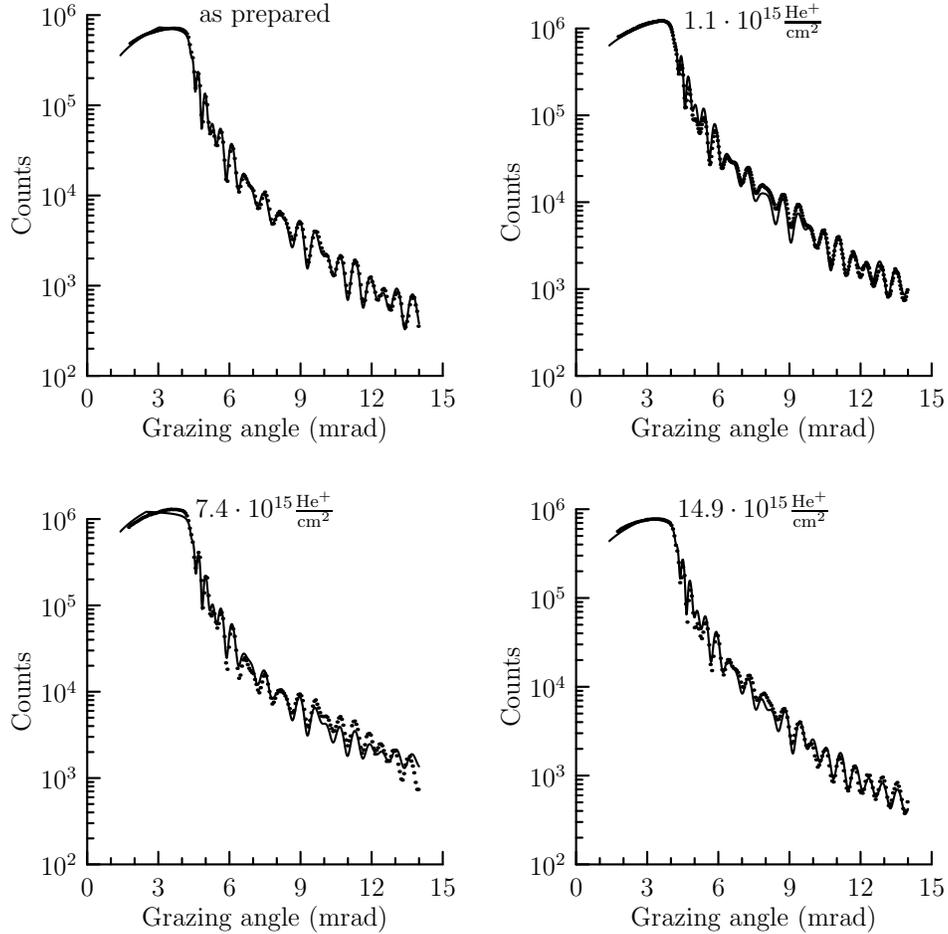} 
\caption{X-ray reflectometry spectra of samples irradiated with different \(\text{He}^+\) doses and the fitted curves obtained with \ProgramName{}.}\label{figureNonResRefl}
\end{figure}

\begin{figure}
\includegraphics[clip,width=125mm]{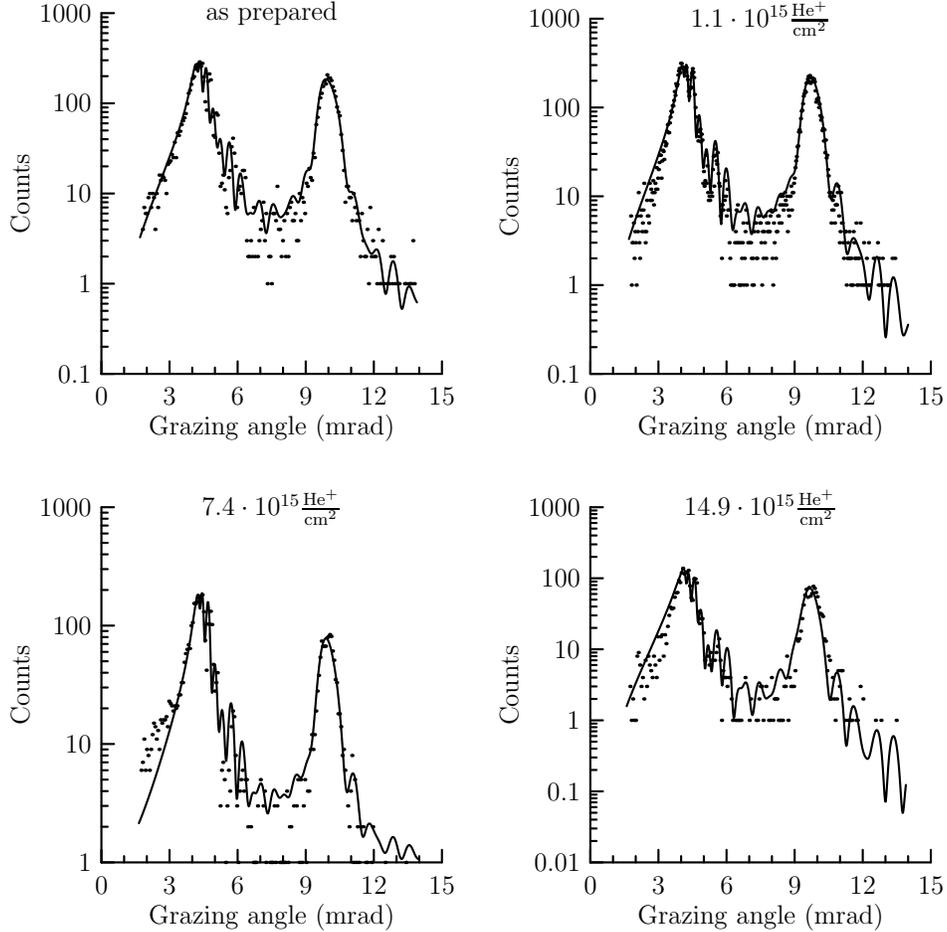} 
\caption{Synchrotron \Mossbauer{} reflectometry in time integrated spectra of samples irradiated with different \(\text{He}^+\) doses and the fitted curves obtained with \ProgramName{}.}\label{figureDiff}
\end{figure}

In the following we will show a few examples used for fitting. These are experiments performed to determine the self-diffusion coefficient of iron atoms in FePd alloys \cite{DaniCikk}. For these experiments isotopic 
Pd(1 nm) \([^{57}\ttinf{Fe}_{47\%}\ttinf{Pd}_{53\%}\)(2 nm) \(^{\ttinf{nat}}\ttinf{Fe}_{47\%}\ttinf{Pd}_{53\%}\)(3 nm)\(]_{10}\) Pd(15 nm) Cr(3 nm) MgO(001) 
multilayers were grown, which thereafter were heated or irradiated with \(\ttinf{He}^+\) ion beam. The effect of the latter treatment can also be modelled as a diffusion of the iron atoms. X-ray reflectograms (Fig. \ref{figureNonResRefl}), and nuclear resonant reflection spectra in time integrated mode (Fig. \ref{figureDiff}) were measured for samples treated with different ion fluxes. For the evaluation of these spectra \ProgramName{} was used with succes. For further details of these experiments and their interpretation see \cite{DaniCikk}.
As it is visible in Figs. \ref{figureNonResRefl}-\ref{figureDiff}, the theoretical results fit well to the experimental data.
\section{Summary}
In this paper we presented a new general extendible program \ProgramName{} for simultaneous simulation and fitting of experimental data of measurements performed on complex systems.
\section{Acknowledgement}
This work was supported by the European Community under the Specific Targeted Research Project Contract No. NMP4-CT-2003-001516 (DYNASYNC). \ProgramName{} was developed in frames of DYNASYNC and it is freely available from \cite{EffiWebPage} with the only provision of proper acknowledgement in future publications.

\appendix

\section{Auxiliary functions needed writting simulation and initialization functions}\label{sec:appA}

Before the list we have to mention some additional facts, which we have to know in order to use them:

\section{Adding a new model type}\label{sec:appB}
In the previous appendix we saw the auxiliary functions needed to write simulation and initialization functions. In this appendix we will see, how we can add a new model type to the program. In future for this task we will use another program with graphical user interface, in order to decrease the number of possible errors in this process. But now we have to write a C++ program, as we will see this can be done quite mechanically.

Here we will show only the most important steps, things, tricks, but not all of them. In the following, short description parts, explanations will precede the corresponding code fragments.


\begin{thebibliography}{00}
\bibitem{Spiering00} H.~Spiering, L.~De\'{a}k, L.~Botty\'{a}n, Hyperfine Interact. 125, 197, (2000)
\bibitem{NDL00} D.L.~Nagy, L.~Botty\'{a}n, L.~De\'{a}k, E.~Szil\'{a}gyi, H.~Spiering, J.~Dekoster, G.~Langouche, Hyperfine Interact. 126, 353, (2000)
\bibitem{Kulcsar71} K.~Kulcs\'{a}r, D.L.~Nagy, L.~P\'{o}cs, in: {\it Proc. Conf. on M\"{o}ssbauer Spectrometry}, Dresden (1971). 
\bibitem{Muller82} E.W.~M\"{u}ller, MOSFUN, Laboratory report, Anorganische Chemie und Analytische Chemie, Johannes Gutenberg-Universit\"{a}t , Mainz (1982).
\bibitem{EffiWebPage} http://www.fs.kfki.hu
\bibitem{IFEFFIT}M.~Newville, J. Synchrotron Rad. 8, 322-324, (2001).
\bibitem{RefFIT}RefFIT written by A.~Kuzmenko is available at http://optics.unige.ch/alexey/reffit.html
\bibitem{SIMULREFLEC} SimulReflec written by F. Ott and coworkers is available at http://www-llb.cea.fr/prism/programs/simulreflec/simulreflec.html
\bibitem{DeakOffsp} L.~De\'{a}k, L.~Botty\'{a}n, D.L.~Nagy, H.~Spiering, Yu.N.~Khaidukov, Y.~Yoda sent to journal, available at http://aps.arxiv.org/abs/0709.2763
\bibitem{NumRec} W.H.~Press, S.A.~Teukolsky, W.T.~Vetterling, B.P.~Flannery Numerical Recipes in C (Fortran), Cambridge University Press
\bibitem{DaniCikk} D.G.~Merkel, et al. to be published 




\end{thebibliography}
\end{document}